\documentclass[12pt,preprint]{aastex}

\usepackage{graphicx}

\newcommand{\beq}{\begin{equation}}
\newcommand{\eeq}{\end{equation}}
\newcommand{\beqa}{\begin{eqnarray}}
\newcommand{\eeqa}{\end{eqnarray}}

\def\la{\lower.5ex\hbox{$\; \buildrel < \over \sim \;$}}
\def\ga{\lower.5ex\hbox{$\; \buildrel > \over \sim \;$}}

\begin{document}

\title{Accretion History of the  Milky Way Dark Matter Halo\\  and the Origin of its Angular Momentum}

\author{P.~J.~E. Peebles}  
\affil{Joseph Henry Laboratories, Princeton University, Princeton, NJ 08544}

\begin{abstract}
The flow of dark matter into the Milky Way and Large Magellanic Cloud in a numerical solution for the evolution of the gravitational field of the neighboring galaxies yields a growth history for the dark matter halo of the Milky Way that ends up with angular momentum roughly in the observed direction, and it produces a dark matter stream around the Large Magellanic Cloud that resembles the Magellanic Stream of atomic hydrogen.
\end{abstract}
\maketitle

\section{Introduction}\label{sec:1}
It is natural to ask whether a dynamical solution for the evolution of relative positions of the nearby galaxies produces a gravitational field that can give a useful approximation to the formation of the dark matter (DM) halo of the Milky Way (MW) galaxy, and possibly account for its angular momentum. This analysis uses a dynamical solution constrained by measurements of galaxy distances and motions in and near the Local Group (LG), along with the condition that peculiar velocities are growing at high redshift (Peebles \& Tully 2013b, hereafter PT2). In this PT2 solution the galaxies, or dynamical actors, are approximated as rigid distributed masses that interact only by gravity. I present the result of replacing the actors representing the MW and the Large Magellanic Cloud (LMC) by halos of dark matter  particles. In five trials this reconstruction produces present angular momentum of the MW DM halo that is directed at galactic latitudes in the range $b=-20^\circ$ to $-80^\circ$ (Sec.~\ref{sec:am}). That is, the reconstruction cannot predict the direction of rotation of the MW, but with significant probability it puts the DM halo angular momentum in the general direction of what is observed, $b=-90^\circ$. The reconstructed MW halo evolves through a quite elongated shape at redshift $z\sim 3$, and assembly is close to complete at $z=1$ (Sec.~\ref{sec:Evolution}). This evolution might be a useful guide to a simulation of the formation of the stellar disk. The reconstructed halo of the LMC develops a DM stream that looks reasonably similar to the Magellanic Stream of  atomic hydrogen. The leading arm of the DM stream is shorter than the trailing part, but the difference  is not as pronounced as in atomic hydrogen. That is, the Magellanic Stream could have developed out of the gravitational interactions modeled here, but the picture still requires the ram pressure of a corona, or perhaps a more diffuse and  more highly ionized leading stream (Sec.~\ref{sec:MS}). Section~\ref{Sec:discussion} offers thoughts on how this study of halo reconstruction might be improved.

\section{The Reconstruction}

\subsection{Dynamical Actors}\label{Sec:generalities}

This computation uses the PT2 solution that best fits 85 parameters for 15 LG galaxies and 4 massive external actors (including measured present distances, redshifts, proper motions, masses from K-band luminosities, and the MW circular velocity, along with the constraint that the initial peculiar velocities are growing). In this solution 6 of the 85 parameters are outside three nominal standard deviations. This would be too many if the model and standard deviations were accurate, but the model is at best a rough approximation and one must expect that some of the standard deviations are inaccurate. The fit of 79 of the 85 parameters within $3\sigma$  argues that the model may be close enough to reality to make it an interesting basis for a reconstruction of the evolution of the MW and LMC DM halos.

The starting assumption in the dynamical model is that the mean motion of the DM that is merging to form a galaxy or concentration of galaxies is usefully represented by the motion of a mass tracer, a dynamical actor that moves under the gravitational attraction of neighboring systems represented by their dynamical actors. The dynamical actors are approximated as rigid bodies (that is, fixed physical mass distributions). This is unrealistic but likely to be seriously problematic only when galaxies merge, but the mass tracer is supposed to be following the effective center of the mass that ends up in the present-day actor, including merging. The plot of orbits of DM particles and actors in Section~\ref{Sec:orbits} illustrates the idea.

The constraint that the actors have growing initial peculiar velocities is meant to approximate the condition that the growing mode in linear perturbation theory dominates the departure from homogeneity at high redshift. The initial velocities of the actors are computed from mixed boundary conditions ---  present positions and initially growing peculiar velocities --- using the numerical action method (Peebles, Phelps, Shaya, \& Tully 2001 and references therein). The approximation to the behavior of the growing mode in linear theory is illustrated in Figure~1 in Peebles \& Tully (2013a, PT1). The behavior looks reasonable, but it is an approximation  that requires correction to the halo particle initial conditions computed in linear theory, as discussed next. 

\subsection{Computation of Halo Particle Orbits}\label{Sec:numerical methods}

The orbits of the halo particles are computed by numerical integration of the equation of motion computed forward in time from initial positions based on the PT2 initial positions of the actors (which are derived from the mixed boundary conditions of present positions and initially growing velocities). The equation of motion in coordinates comoving with the general expansion of the universe in the standard relativistic $\Lambda$CDM cosmology is 
\beq
{d\over dt}a^2{d\vec x_i\over dt}={\vec g_i\over a}.
\label{eq:ofmotion}
\eeq
The physical peculiar velocity is $a\,d\vec x_i/dt$ and the physical peculiar acceleration is $\vec g_i/a(t)^2$ where, ignoring the truncation of the inverse square law,
\beq
\vec g_i=\sum{Gm_j\vec x_{ji}\over x_{ji}^3} + {1\over 2}\Omega_m H_o^2a_o^3 \vec x_{i}.\label{eq:acceleration}
\eeq
The origin of coordinates is the center of mass of all the particles in the model (as discussed in Peebles 1994, eq.~[7]), $a_o$ is the present value of the expansion parameter $a(t)$, Hubble's constant is $H_o$, the matter density parameter is $\Omega_m$ (where in this computation $H_o=70\hbox{ km s}^{-1}\hbox{ Mpc}^{-1}$ and $\Omega_m=0.27$), the cosmological constant is represented by $\Omega_\Lambda =1-\Omega_m$, and space curvature and the mass density in radiation are taken to be negligibly small. For the halo particles, $x_{ji}^3$ in equation~(\ref{eq:acceleration}) is replaced by $x_{ji}^3 + x_{\rm in}^3$, where the physical cutoff length is $a(t)x_{\rm in}=10$~kpc. The mass distributions in the actors are modeled as the truncated limiting isothermal gas spheres in equation~(2) in PT1.

In linear perturbation theory at high redshift, $\vec g_i\propto a(t)\propto t^{2/3}$ (because the effect of radiation at very high redshift is ignored), and equation~(\ref{eq:ofmotion}) integrates to 
\beq
\vec x_i(t_2) = \vec x_i(t_1) + {3\vec g_i(t_1)t_1^2(a(t_2) - a(t_1))\over 2a(t_1)^4}, \label{eq:initvel}
\eeq
at the first two time steps, $t_1$ and $t_2$. This expression must be adjusted, however, because PT2 only approximates linear theory at high redshift. To arrange that the halos arrive acceptably close to the present PT2 positions of MW and LMC, equation~(\ref{eq:initvel}) is changed to 
\beq
\vec x_i(t_2) = \vec x_i(t_1) + {3\vec g_i(t_1)t_1^2(a(t_2) - a(t_1))\over 2a(t_1)^4} + 
	{\vec v_{\rm init}\over a(t_1)}(t_2 - t_1). \label{eq:newinitvel}
\eeq
The initial uniform physical peculiar velocity $\vec v_{\rm init}$ is adjusted separately for the two halos. The main effect is to make the initial halo motions agree with the initial conditions of the MW and LMC actors, which are not exactly moving in accordance with linear theory. The noise from random placing of the halo particles requires an additional $\sim 10$\,\% adjustment of the $\vec v_{\rm init}$. 

Given the initial halo particle position $\vec x_i(t_1)$, equation~(\ref{eq:newinitvel}) gives the position $\vec x_i(t_2)$ at the next step. These serve as initial conditions for leapfrog numerical integration forward in time (as in eq.~[9] in Peebles 1995). The orbits of halo particles and actors are represented in this study by 500 time steps uniformly spaced in $a(t)$ over the expansion of the universe by a factor of ten, that is, starting from redshift $z_i=9$. The mass $1.52\times 10^{12} m_\odot$ of the MW actor in the PT2 solution is converted to 4000 halo particles of mass $3.8\times 10^8 m_\odot$. The choice of 2263 particles for the LMC halo makes its halo particle mass one tenth that of a MW particle. The relatively modest number of particles is limited for convenience of computation by an iMAC desktop computer with a na\" ive particle accelerator. Convenience also dictates ignoring the effect of the mass rearrangement within MW and LMC on the orbits of the remaining actors. The test in Section~\ref{Sec:fwdint} indicates that this is a reasonably good approximation.

\subsection{Reconstruction Method} \label{Sec:reconstruction}

Replacing the MW and LMC actors by DM halos must deal with the sensitivity of present positions to initial conditions. The adopted procedure is to decrease the mass of the  actor in steps, $m_{\rm a}\rightarrow m_{\rm a} - m_{\rm p}$, as each DM particle of mass $m_p$ is added to the halo. The mass $m_{\rm a}$ remaining in the actor defines the comoving radius
\beq
a_o\,x_g = \left(2Gm_{\rm a}\over \Omega_m H_o^2\right)^{1/3}.  
\eeq
A sphere of this radius in a homogeneous universe contains mass $m_{\rm a}$. If this mass were gathered at the center then a test particle placed at the edge of the sphere with zero peculiar velocity would drift away from the concentration with the general expansion of the universe. In the reconstruction the newly added halo particle is placed with homogeneous probability density in the range of radii 
\beq
0.25 x_g< x <1.5 x_g\label{eq:bounds}
\eeq
around the initial position of the actor in the PT2 solution. The upper limit allows the actor to gather matter from distances larger than the initial value of $x_g$ from some directions.  The lower bound is required because at smaller distances particles are overly strongly attracted to the residual mass $m_{\rm a}$ of the actor, making the linear perturbation approximation in equation~(\ref{eq:initvel}) quite inadequate.

After each halo particle is placed its motion is computed under its gravitational interaction with the actors moving in their PT2 orbits, the residual mass in the actor being reconstructed, and the already computed orbits of the previously placed halo particles. The placement of the new halo particle is accepted if it ends up within a target distance $D$ from the present position of the actor, with $D=100$~kpc for MW and 50~kpc for LMC. If the particle ends up further away the initial position is placed again with new random numbers. This reassignment does not significantly slow reconstruction of MW, but it sometimes is a problem for LMC, so after 100 failures to place a new halo particle the LMC reconstruction starts anew. After initial positions for all the halo particles are placed, orbits are computed for all halo particles under their gravitational interaction with each  other and the remaining actors. The halos end up more broadly distributed than $D$ because halo particles are disturbed by the later placing of more particles. 

Linear theory can only be a rough approximation to the halo particle initial velocities because in this computation the initial densities within the reconstructed halos are about twice the cosmic mean. But that error is largely compensated by the condition on $D$ that the halo particles end up in reasonably realistic concentrations. 

After reconstruction the initial velocities $\vec v_{\rm init}$ in equation~(\ref{eq:newinitvel}) are adjusted, largely by hand, to bring the present positions of the peak densities in the halos acceptably close to the PT2 positions. The convenient approximation to the peak position is the central position of a sphere of radius $r$ at which the center of mass of the particles in the sphere is at the center of the sphere. The sphere radii are
\beq
r=30\hbox{ kpc for MW},\qquad r=10\hbox{ kpc for LMC}. \label{eq:averagingspheres}
\eeq
Visual inspection indicates that this approximation satisfactorily locates peak halo densities.

\begin{figure}[t]
\begin{center}
\includegraphics[angle=0,width=6.in]{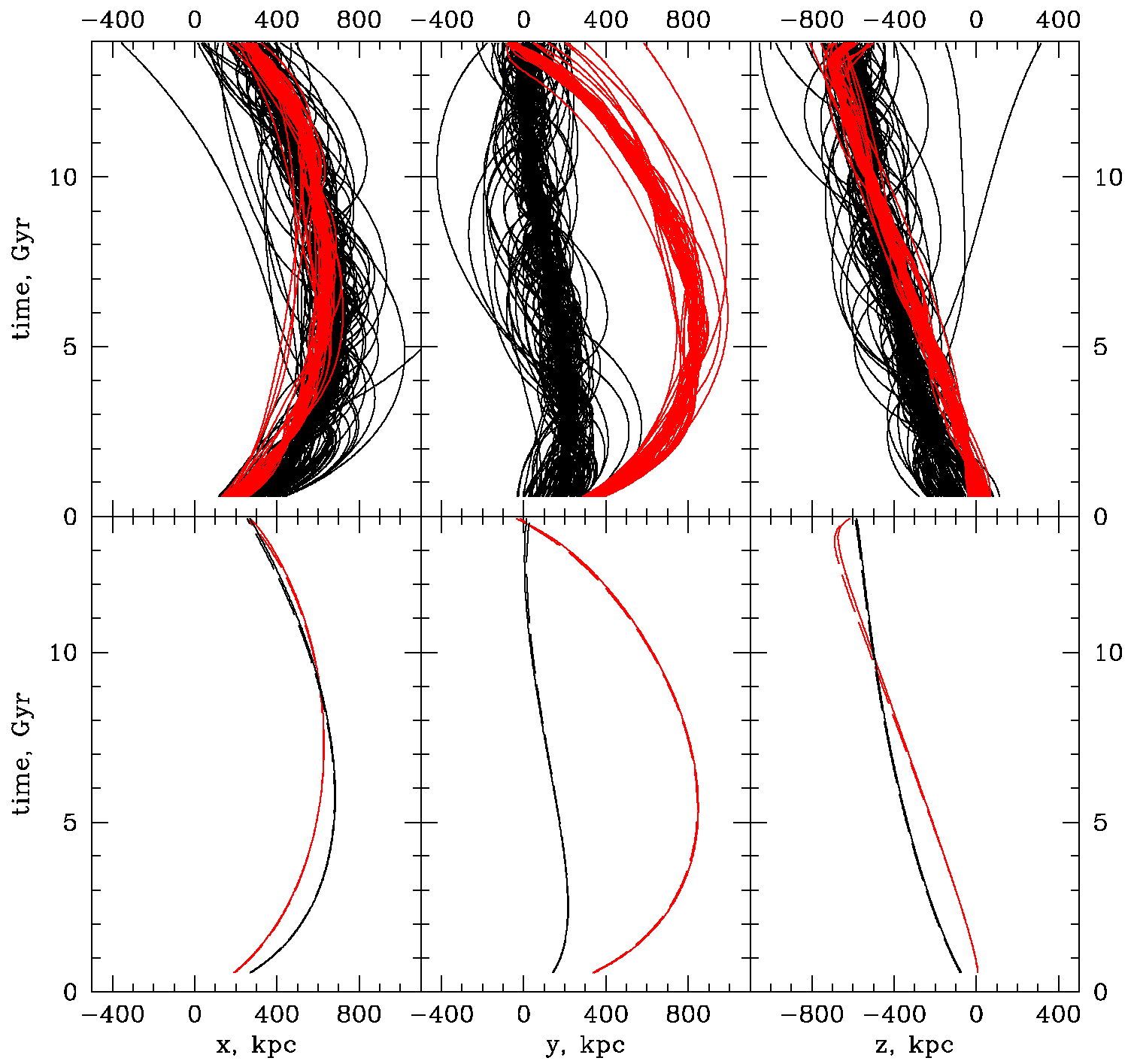} 
\caption{\label{Fig:orbits} \footnotesize Orbits of MW (black) and LMC (red) halo particles. Lengths and times are physical. Coordinates are galactic with origin at the center of mass of the PT2 solution. Upper panels are orbits of samples of 100 MW halo particles and 50 LMC particles. Lower panels compare PT2 actor orbits (solid lines) to the orbits of the centers of mass of the halos (dashed lines).}
\end{center}
\end{figure}

\section{Results}\label{Sec:results}

Five reconstructions with different random numbers are labeled (a) through (e). Where the differences among reconstructions look  small I show case (a), which has angular momentum closest to the MW disk of stars.

\subsection{Halo Particle Orbits}\label{Sec:orbits}

The upper panels in Figure~1 show orbits of samples of 100 of the 4000 MW halo particles and 50 of the 2263 LMC particles in reconstruction (a). Lengths and times are physical and the coordinates are galactic. The origin is the center of mass of the 19 actors in the PT2 solution (which differs from the center of mass of the two halos and the remaining 17 actors by less than 2\,kpc). The lower panels compare the orbits of the centers of mass of the LMC and MW halos, plotted as dashed lines, to the PT2 orbits of the actors, plotted as solid lines. The difference between the orbit of the actor and the center of mass of all the particles belonging to the halo is largely a result of the difference between the latter and the position of the peak halo density. The initial velocities have been adjusted to place the present peak density acceptably close to the present PT2 location of the actor.

In reconstruction (a) the initial proper physical velocities (added to the linear theory velocity in eq.~[\ref{eq:newinitvel}]) are 
\beq
\vec v_{\rm init~MW} =  (110,\,50, -23)\hbox{ km s}^{-1}, \quad 
\vec v_{\rm init~LMC} = (73,\, 146, \, 8)\hbox{ km s}^{-1}.
 \label{eq:newinitvela}
\eeq
The offsets of the present positions of the halo peak densities from the positions of the PT2 actors are
\beqa
&&  \vec\delta_{\rm MW}= \vec r_{\rm MW~ halo} - \vec r_{\rm MW~ actor} = 16\hbox{ kpc to } \ell = 59^\circ,\ b=-46^\circ, \cr
&& \vec\delta_{\rm LMC-MW}= \vec r_{\rm LMC~ halo} - \vec r_{\rm MW~ halo} - \vec r_{\rm LMC~ actor}  +
\vec r_{\rm MW~ actor}  \cr
&&\hspace{5cm} = 8\hbox{ kpc to } \ell = 238^\circ,\ b=-25^\circ.
\label{eq:offsets}
\eeqa
In the first line, $\vec\delta_{\rm MW}$ is the position of the peak of the MW halo relative to the PT2 position of the actor. The offset $\vec\delta_{\rm LMC-MW}$ in the second line compares the position of the LMC halo peak relative to the MW halo peak to the position of the LMC actor relative to the MW actor. Reducing $|\vec\delta_{\rm LMC-MW}|$ well below 10\,kpc is difficult because $\vec\delta_{\rm LMC-MW}$ is a sensitive and complicated function of the $\vec v_{\rm init}$. The offset $\vec\delta_{\rm MW}$ is readily reduced by adjusting $\vec v_{\rm init\ MW}$, but that complicates placing the LMC halo. In the five reconstructions $\vec\delta_{\rm MW}$ ranges from 3~kpc to 35~kpc, and $\vec\delta_{\rm LMC-MW}$ is less than 10~kpc in all reconstructions, which seems adequate for the purpose of this study. The spreads of values of the components of $\vec v_{\rm init~MW}$ and $\vec v_{\rm init~LMC}-\vec v_{\rm init~ MW}$ in the realizations are less than about 10~km~s$^{-1}$, except for the LMC $y$-components, which spread over $25$~km~s$^{-1}$. 

\begin{figure}[t]
\begin{center}
\includegraphics[angle=0,width=6in]{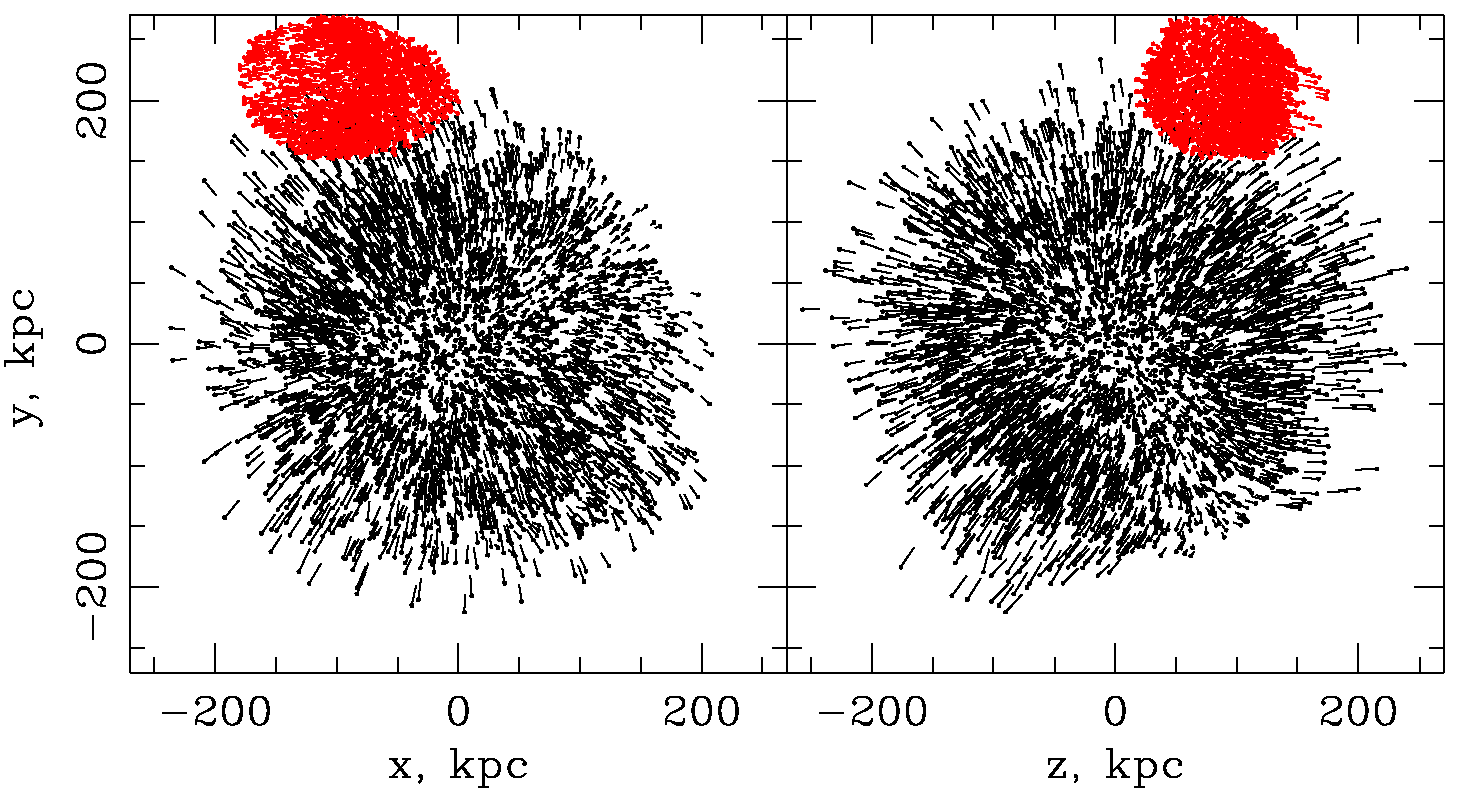} 
\caption{\label{Fig:z9} \footnotesize Initial positions and peculiar velocities of the halo particles projected onto the present plane of the Milky Way in the left-hand panel, and in an orthogonal projection in the right-hand panel..}
\end{center}
\end{figure}

\subsection{Initial Conditions}\label{sec:InitialConditions}

Figure~\ref{Fig:z9} shows initial positions and velocities of the MW (black) and LMC (red) halo particles in reconstruction (a) in Figure~1. The circles mark the halo particle initial positions relative to the initial position of the MW actor. The lines point in the direction of the initial peculiar velocity relative to the initial peculiar velocity of the MW actor, projected onto the plane, and the line lengths are proportional to the magnitude of the projected velocity. 

Near the edge of the initial distribution of MW halo particles, at physical radius 200~kpc, the outward Hubble flow is 230~km~s$^{-1}$, and the average inward peculiar velocity at this radius is about $\sim 100$~km~s$^{-1}$, meaning the halo is initially expanding at about half the rate of general expansion of the universe. This means that, as previously noted, the use of linear theory for initial velocities is a crude approximation. 

The projected initial peculiar velocities near the center of the MW halo in Figure~\ref{Fig:z9} are close to zero, a result of two effects. First, the streaming velocites $\vec v_{\rm init}$ (eq.~[\ref{eq:newinitvela}]) applied to the halo particle initial conditions compensate for the departure of the initial behavior of the actors from linear theory, making the halo move with the actor. Second, the initial peculiar velocities in the outer parts of the MW halo are close to radial. A measure of the present tendency to radial orbits is discussed in Section~\ref{Sec:MWvc}. 

Figure~\ref{Fig:z9} shows a generally subdominant coherent pattern of transverse streaming motions caused by the gravitational field of the external actors along with the interaction between the MW and LMC halos. The nonradial pattern signifies transfer of angular momentum to the halo. 
 
 \begin{figure}[t]
\begin{center}
\includegraphics[angle=0,width=6.in]{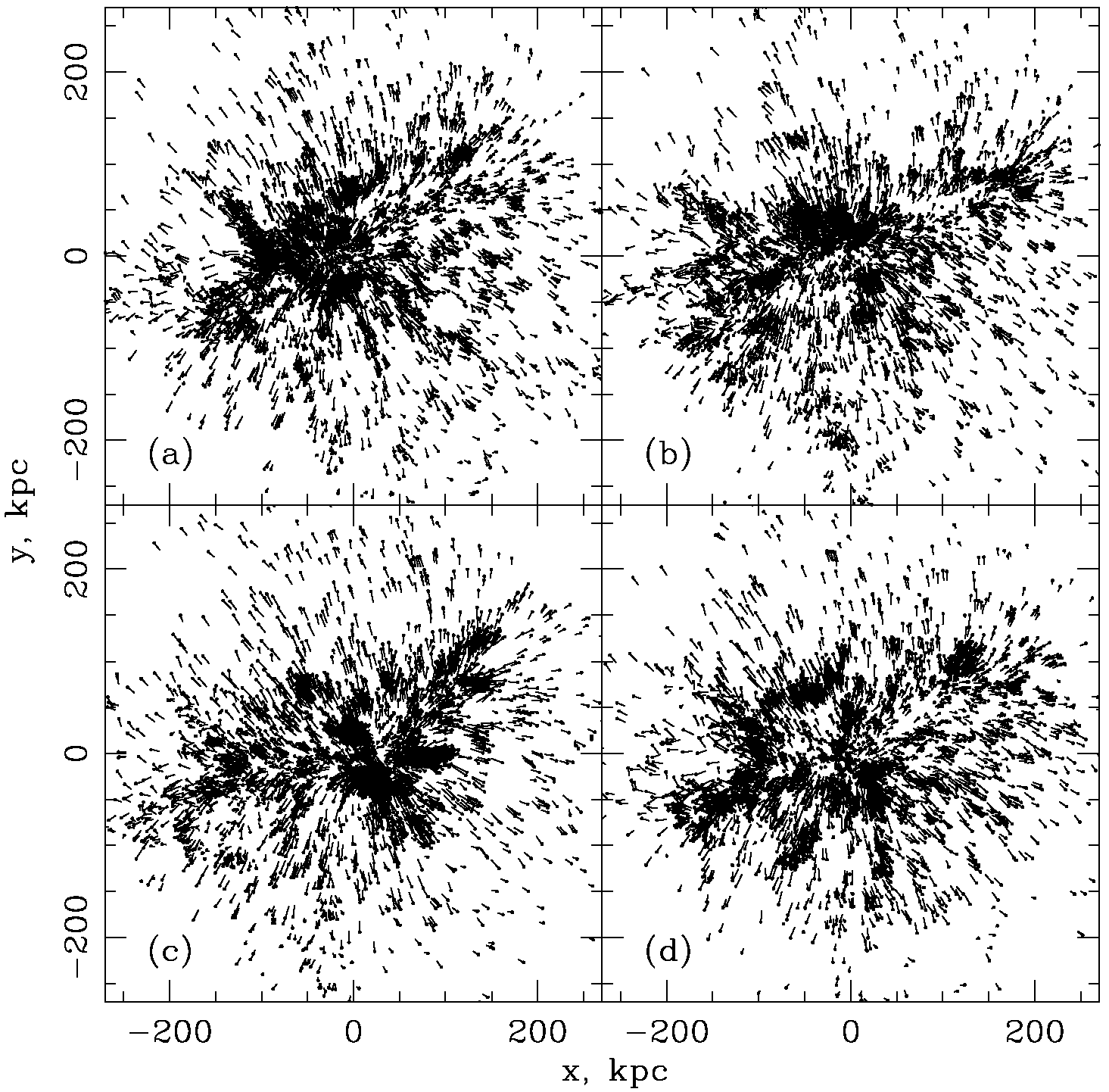} 
\caption{\label{Fig:z4} \footnotesize Halo particle positions and peculiar velocities projected on the present WM plane at redshift $z=4$ in four of the reconstructions.}
\end{center}
\end{figure}

\begin{figure}[t]
\begin{center}
\includegraphics[angle=0,width=6.in]{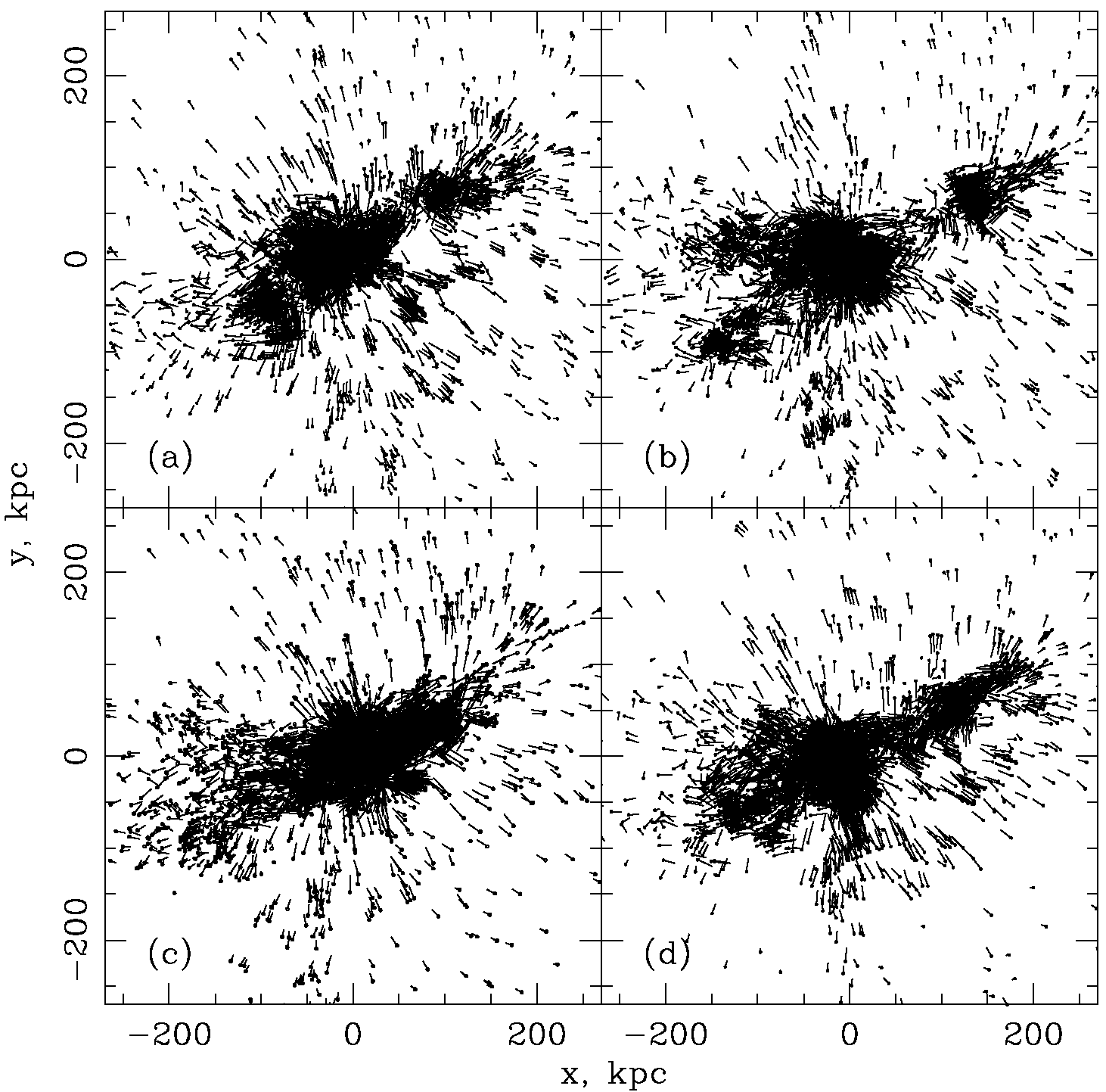} 
\caption{\label{Fig:z3} \footnotesize Halo particle positions and peculiar velocities at redshift $z=3$.}
\end{center}
\end{figure}

\subsection{Evolution of the MW Halo}\label{sec:Evolution}

The initially close to homogeneous and isotropic halo in Figure~\ref{Fig:z9} causes the initial expansion and early stages of contraction of the halos to be artificially close to uniform, as one sees in the upper panels of Figure~\ref{Fig:orbits}. After $t\sim 2.5$~Gyr, redshift $z\sim 3$, there is little indication of evolution in the braided near stationary appearance of the halo particle orbits. Figure~\ref{Fig:z4} shows the situation in the MW halo somewhat before this, at $z=4$, in plots of particle positions and velocities relative to the MW actor position and velocity at that time, and projected onto the present plane of the MW. Four reconstructions are labeled in the four panels. The distinct differences among the halo particle distributions at $z=4$ show the effect of shot noise in the initial positions. The less pronounced similarities in the patterns of positions and velocities show the effect of the gravitational field imposed by the actors, along with the effect of the interaction between the two reconstructed halos.

Figure~\ref{Fig:z3} shows the situation at $z=3$, when the braided orbits have started to look stationary, and one might say that the reconstructed halos have developed pronounced ``subhalos.'' At this redshift one sees clear similarities of the patterns in the halo reconstructions.  Maps at $z=3$ in the $x$-$z$ plane give a similar impression. The PT2 solution predicts a considerably elongated WM DM halo at $z=3$. 

\begin{figure}[t]
\begin{center}
\includegraphics[angle=0,width=6.in]{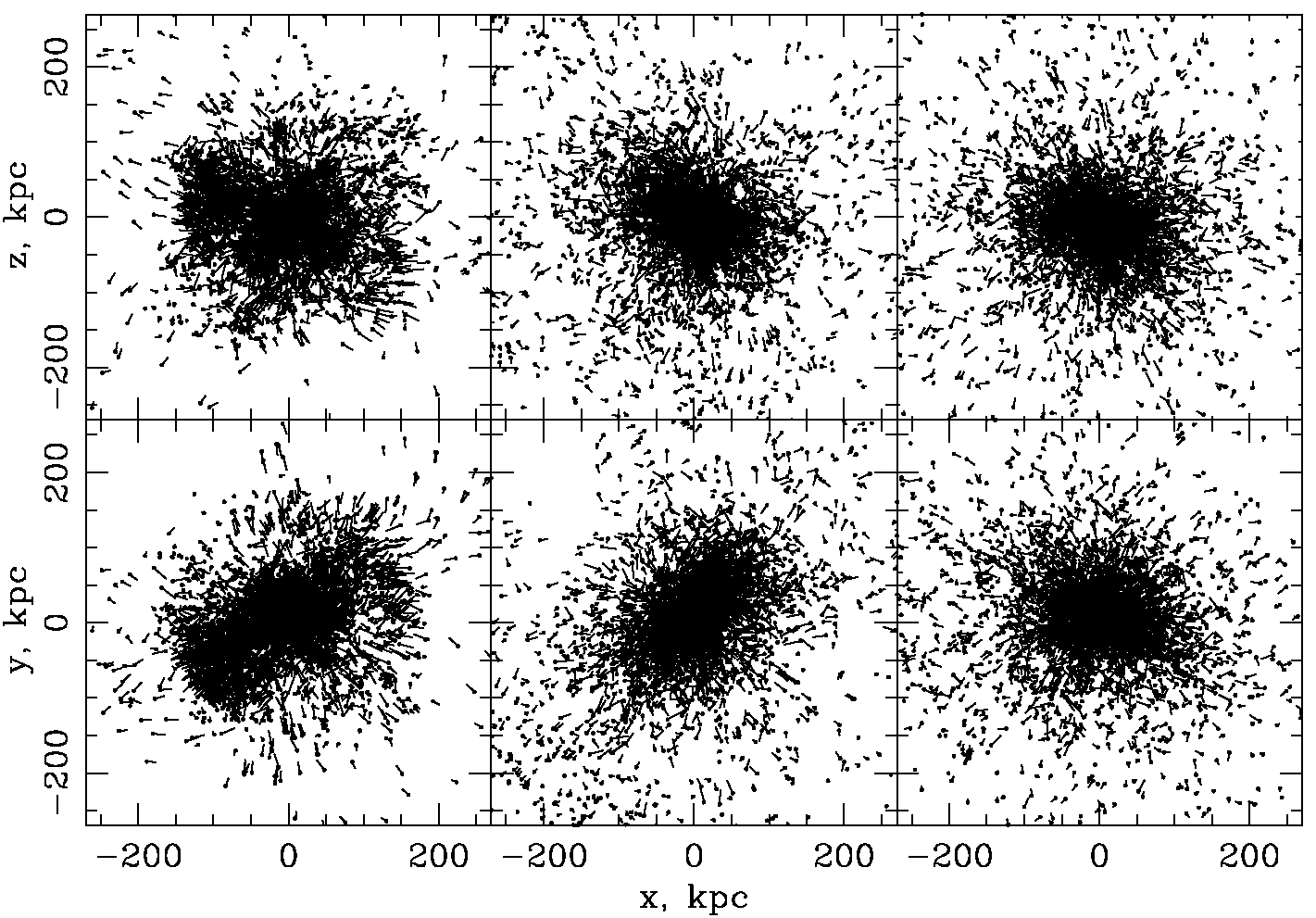} 
\caption{\label{Fig:z210} \footnotesize MW Halo particle positions and peculiar velocities in reconstruction~(a) at redshifts $z=2$, 1, and 0 from left to right, in orthogonal projections. For clarity the LMC halo particles are not plotted. }
\end{center}
\end{figure}
Figure~\ref{Fig:z210} shows the situation in reconstruction~(a) at lower redshifts and in two projections. At $z=2$, in the left-hand panels, there still is some transient substructure, and at $r\sim 150$~kpc particles are still streaming toward the halo center, but comparison with the maps at $z=1$ in the central panels and $z=0$ in the right-hand panels indicates that at $z\la 2$ the mean radial distribution of halo particles is close to statistical equilibrium. This is the same impression offered by the near stationary width of the braided orbits in Figure~\ref{Fig:orbits}  at $z<3$. At $z=1$, in the middle panels in Figure~\ref{Fig:z210}, the halo has relaxed to a close approximation to an ellipsoid, with little noticeable pattern in the halo particle velocities, and  the orientation has remembered the alignment of substructure in Figure~\ref{Fig:z3}. At $z=0$, in the right-hand panels, the spheroid has rotated and the eccentricity is smaller, the latter possibly a consequence of artificial relaxation of the relatively small number of halo particles. The halo evolution looks similar to Figure~\ref{Fig:z210} in the other reconstructions, though the position angles at $z=0$ are noticeably different.  

\begin{figure}[t]
\begin{center}
\includegraphics[angle=0,width=3.5in]{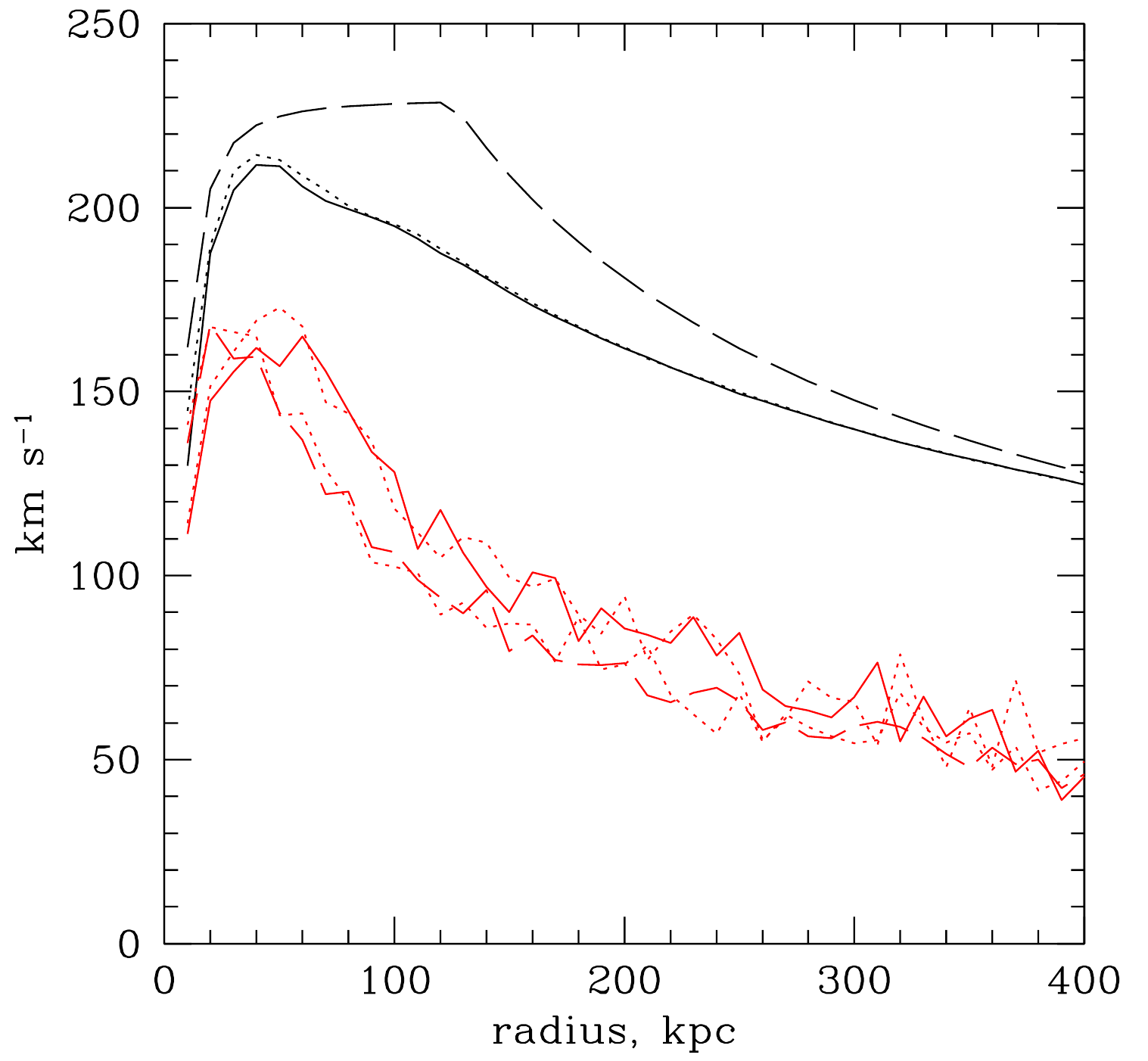} 
\caption{\label{Fig:vc} \footnotesize Circular velocities in the MW halo and the PT2 actor model (solid and dashed black lines), and velocity dispersions in the radial and transverse directions (solid and dashed red lines) in reconstruction (a). The dotted lines show a test of sensitivity to the disturbances of the other actors (Sec.~\ref{Sec:fwdint}).}
\end{center}
\end{figure}

\subsection{Present Structure of the Simulated MW Halo}\label{Sec:MWvc}

Figure~\ref{Fig:vc} shows measures of the present structure of the MW halo reconstruction (a). The dashed black line is the rotation curve of the PT2 model for the mass distribution in the MW actor. The solid black line is the  rotation curve $v_c(r)$ for the circularly averaged mass distribution in the reconstructed halo. The reconstructed halo mass is the same as the MW actor mass, but one sees that the halo mass distribution is  less concentrated at $r\sim 50$ to 300~kpc. The effect of the broader mass distribution on the orbit of the LMC halo has been compensated by adjustment of the initial velocities (eq.~[\ref{eq:newinitvel}]), as a correction added to the effect of the departure of the initial conditions from linear theory.  

The velocity dispersions in the MW halo are plotted as red lines, solid for the radial part $\sigma_r$ and dashed for the transverse component  $\sigma_t$. The radial direction is referred to the peak of the MW halo density and the velocities are relative to the mean velocity of the MW halo. The dispersions are spherically averaged in disjoint 10~kpc radius bins, which makes the noise larger than for $v_c$. In an isotropic velocity distribution  $\sigma_t/\sigma_r=2^{1/2}$. Since $\sigma_t$ is less than $\sigma_r$ the orbits tend to the radial direction. For a rough check of the Jeans equation I have compared the gradient of the number density times $\sigma_r$, in the crude approximation of differences between neighboring 10~kpc radial intervals, to $\sigma_t^2 -2\sigma_r^2 -v_c^2$. At radii 30 to 200 kpc  the two sides of the Jeans equation scatter by about 40~\% in the individual radius bins, with little evidence of systematic differences. Outside this range of radii the noise is too large for a meaningful check. The evidence from this rough test is that the reconstructed MW halo has relaxed to a near steady state. This agrees with the impressions from the braided orbits in Figure~\ref{Fig:orbits} and the maps at low redshifts in Figure~\ref{Fig:z210}. 

\begin{figure}[ht]
\centering
\begin{minipage}[t]{0.49\linewidth}
\includegraphics[angle=0,width=3.in]{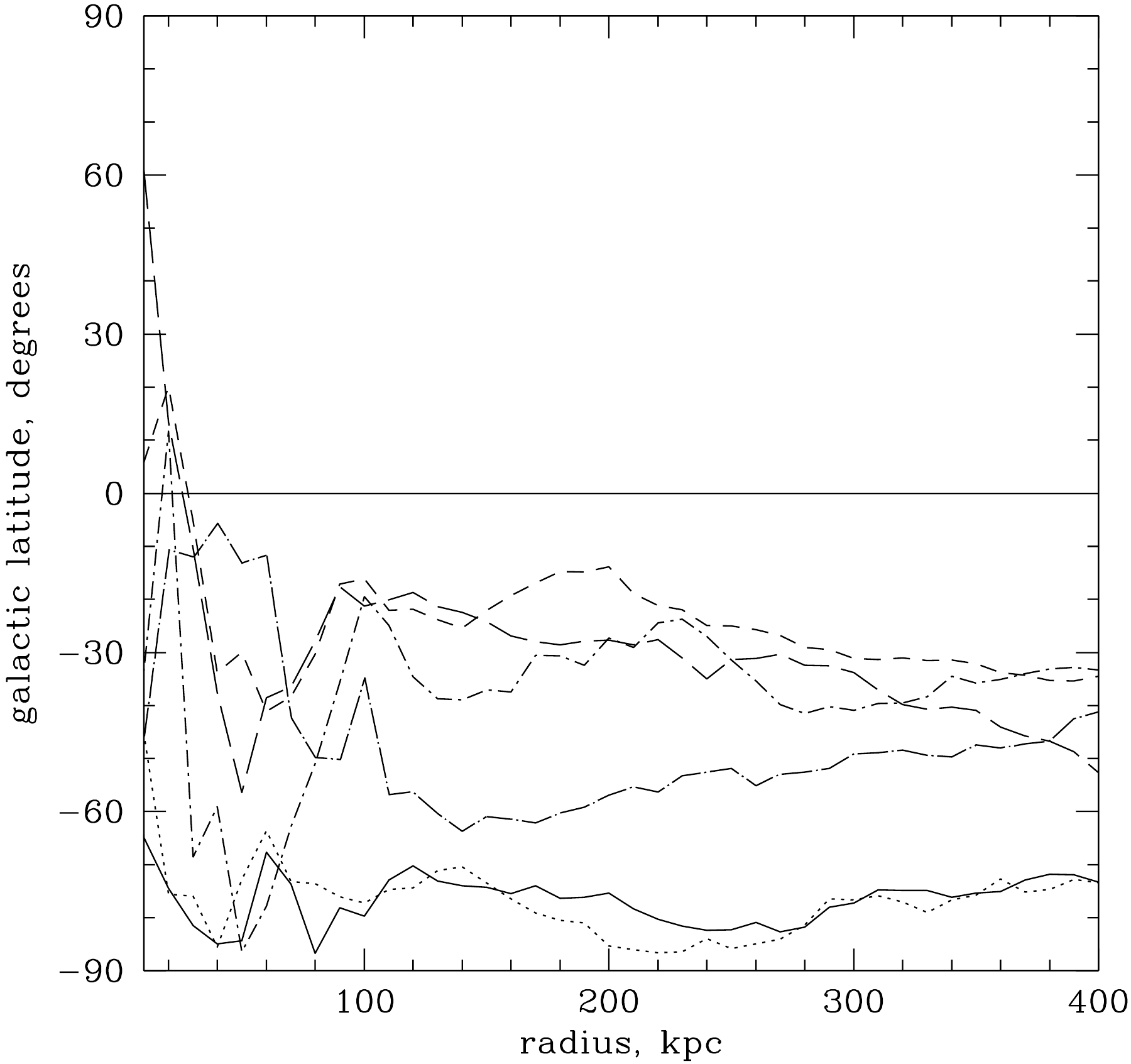}
\end{minipage}
\begin{minipage}[t]{0.49\linewidth}
\includegraphics[width=3.1in]{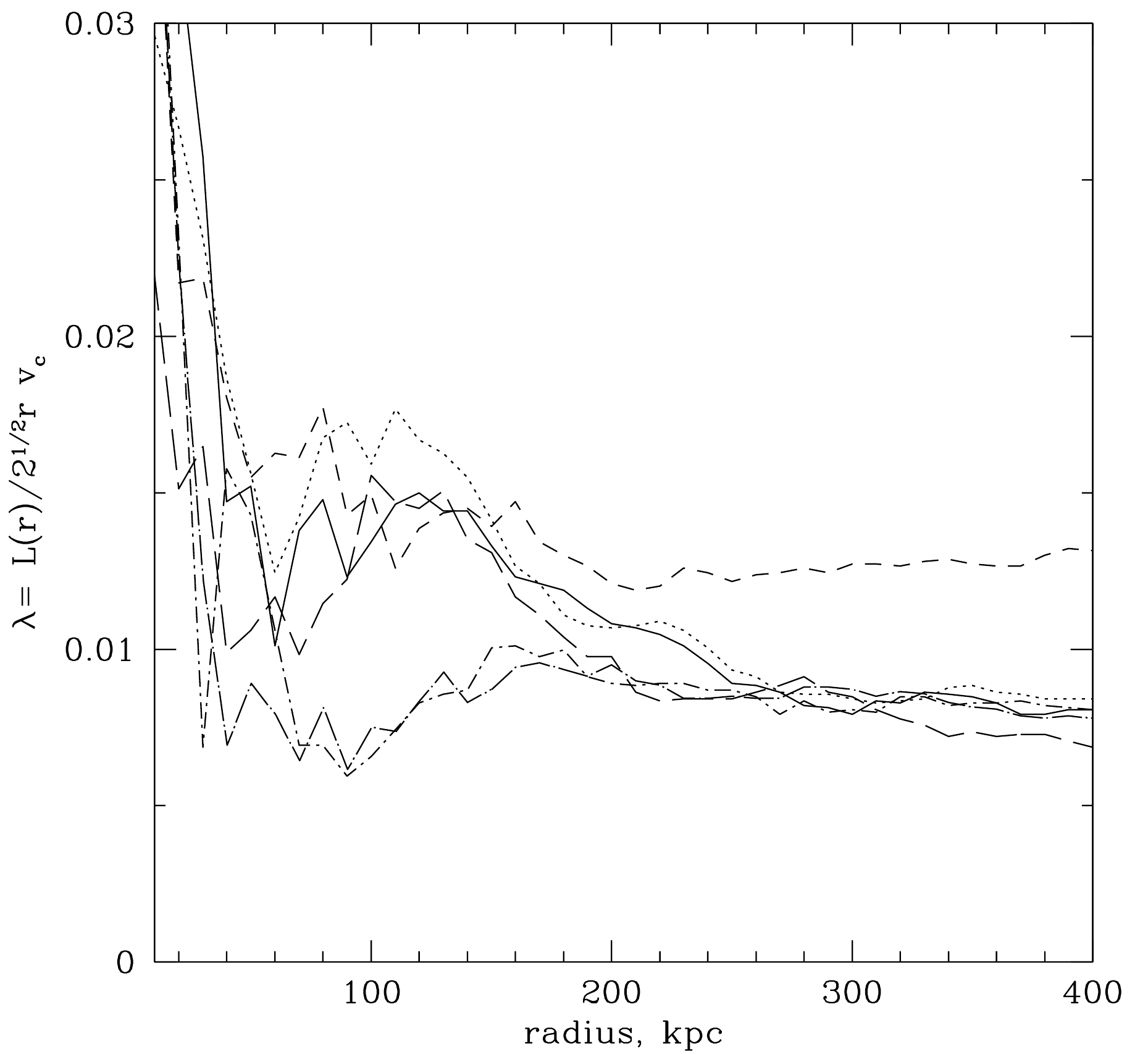}
\end{minipage}
\caption{\footnotesize  Angular momentum of the reconstructed Milky Way halos as a function of limiting radius. The left-hand panel is the galactic latitude of the direction of the angular momentum vector. The right-hand panel is the measure of the magnitude of the angular momentum per unit mass in equation~(\ref{eq:lambda}) The solid line for reconstruction~(a) is to be compared to the dotted line discussed in Section~\ref{Sec:fwdint}.}
\label{Fig:am}
\end{figure}

\subsection{Angular Momentum of the Simulated MW Halo}\label{sec:am}

The angular momentum per unit mass is computed from the $z=0$ particle positions $\vec r_i$ relative to the MW halo density peak and the physical velocities $\vec v_i$ relative to the mean velocity of the MW halo, in the expression
\beq
\vec L = \langle \vec r_i\times\vec v_i\rangle - \langle \vec r_i\rangle\times\langle\vec v_i\rangle.
\eeq
The averages are computed over MW halo particles at radius $|\vec r_i|<r$, as a function of the limiting distance $r$ from the halo center. It would be better to consider the angular momentum in disjoint shells of radius, but the estimates are too noisy.

The left-hand panel in Figure~\ref{Fig:am} shows the galactic latitude of the direction of the angular momentum vector, and the right-hand panel shows the magnitude in the  Bullock, Dekel, Kolatt, {\it et al.} (2001) measure, 
\beq
\lambda(r) = {L(r)\over 2^{1/2}\,r\,v_c(r)},\label{eq:lambda}
\eeq
where $v_c(r)$ is the circular velocity at the limiting radius $r$. The solid black lines show reconstruction (a), which is to be compared to the dotted lines that show the effect of the experiment in Section~\ref{Sec:fwdint}. Reconstructions~(b) to (e) are plotted as long dash, short dash, dot-short dash, and dot-long dash lines. 

At $r< 20$~kpc the noise is too large for meaningful conclusions. At $r=20$~kpc to 400~kpc the direction of 
the angular momentum vector in the five reconstructions varies from $b_{\rm am}\sim -20^\circ$ to $b_{\rm am}\sim -80^\circ$. The substantial scatter is due to the shot noise in the initial positions of the DM halo particles. The angular momentum direction $b_{\rm am}$ in a given realization varies with limiting radius $r$ by a lesser amount than the scatter among the five reconstructions. This relatively slow variation of $b_{\rm am}$ with $r$ may result from mixing by the near radial halo particle orbits, but the noise is too large to check for the expected decrease of $\lambda(r)$ with increasing $r$ in this picture. It is notable that at $r>20$\,kpc all five $b_{\rm am}$ are in the southern galactic hemisphere, where they ought to be if the direction of the MW DM angular momentum is similar to the MW disk of stars. The magnitude of the angular momentum scatters around $\lambda = 0.01$, which is at the low end of the distribution of values found in $\Lambda$CDM N-body simulations (Bullock {\it et al.} 2001; Trowland, Lewis, \& Bland-Hawthorn 2013). This might mean that the MW halo has unusually low angular momentum. Perhaps a more plausible idea is that the model lacks the angular momentum produced by the coupling of clumps of matter in the young halo to the gravitational field of the external actors. 

\begin{figure}[htpb]
\begin{center}
\includegraphics[angle=0,width=5.5in]{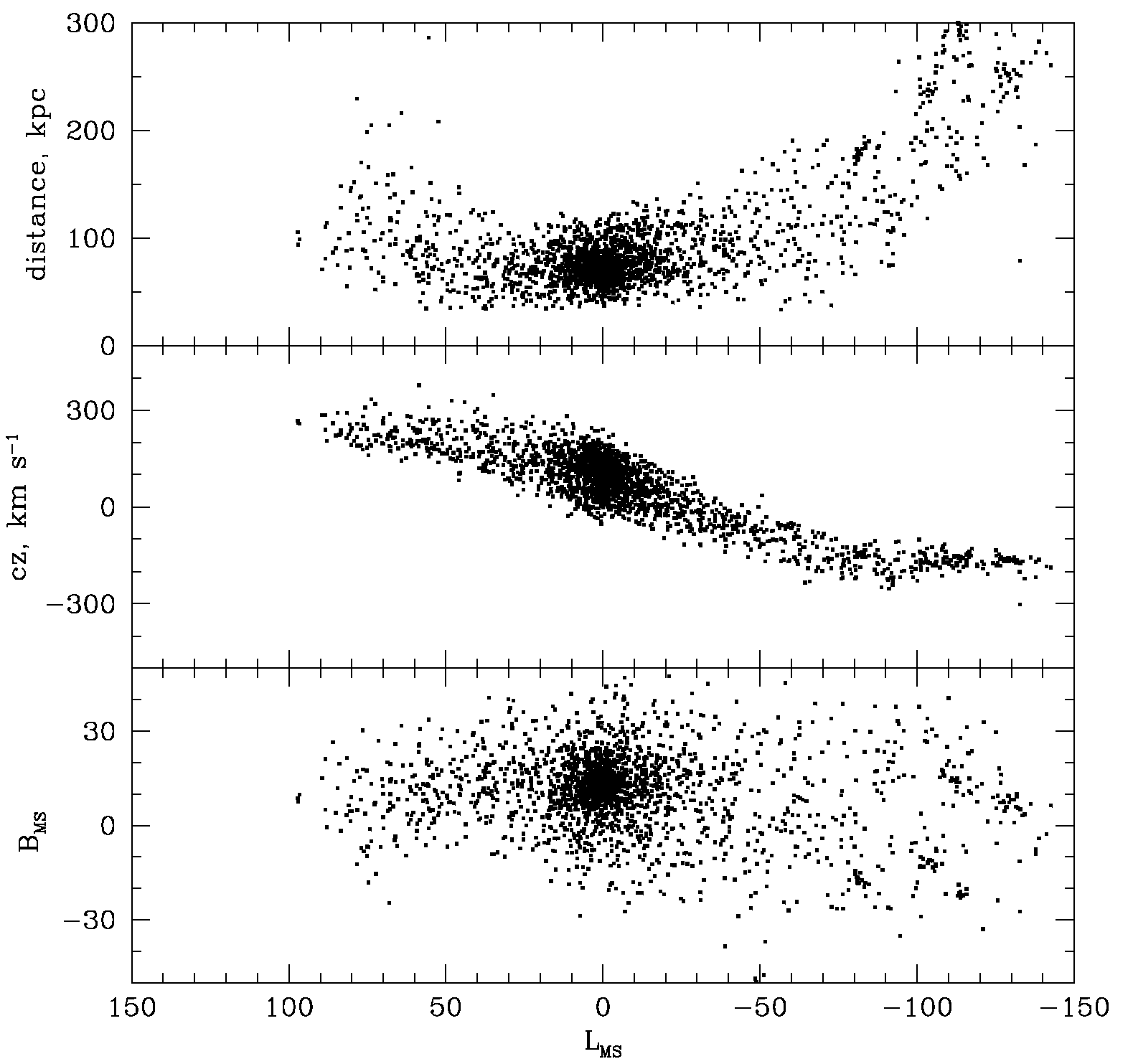} 
\caption{\label{Fig:MS} \footnotesize The LMC dark matter halo particle distribution at $z=0$ in Magellanic Stream coordinates.}
\end{center}
\end{figure}

\subsection{The Magellanic Dark Matter Stream}\label{sec:MS}

Figure~\ref{Fig:MS} shows the present distribution of LMC halo particles in Magellanic Stream coordinates (Nidever, Majewski, \& Butler Burton 2008) in reconstruction~(a). The positions and redshifts are heliocentric, the redshifts based on the MW circular velocity $v_c=230$~km~s$^{-1}$ in the PT2 solution and the Sch{\"o}nrich, Binney, \& Dehnen (2010) Solar Motion. The results resemble the still simpler simulation of the Magellanic Stream in PT2, but in Figure~\ref{Fig:MS} the central concentration is more pronounced and the asymmetry of leading and trailing streams slightly greater. The streams in the other four reconstructions are similar. The HI in the Magellanic leading stream is far less prominent relative the trailing part (Nidever, Majewski, Butler Burton \& Nigra  2010). Perhaps this is an effect of ram pressure by a plasma corona of the Milky Way, or perhaps it reflects a difference of fractional ionization. 

The redshift in the DM leading stream increases with increasing $L_{\rm MS}$, while the HI redshift is closer to constant and at $L_{\rm MS}\sim 50^\circ$ shows a tendency to decrease with increasing $L_{\rm MS}$. In the trailing stream at $L_{\rm MS}\sim -100^\circ$ the redshift in the reconstructed halo is $\sim -250$~km~s$^{-1}$, while the observed HI redshift is $\sim -350$~km~s$^{-1}$. These clear differences between model and measurements may be in part a consequence of the smaller circular velocity in the MW halo reconstruction (Fig.~\ref{Fig:vc}). But the notable and arguably encouraging result is the resemblance of the pattern of redshifts and angular positions in the reconstructed DM stream to the Magellanic HI Stream. 

\begin{figure}[ht]
\begin{center}
\includegraphics[angle=0,width=5.in]{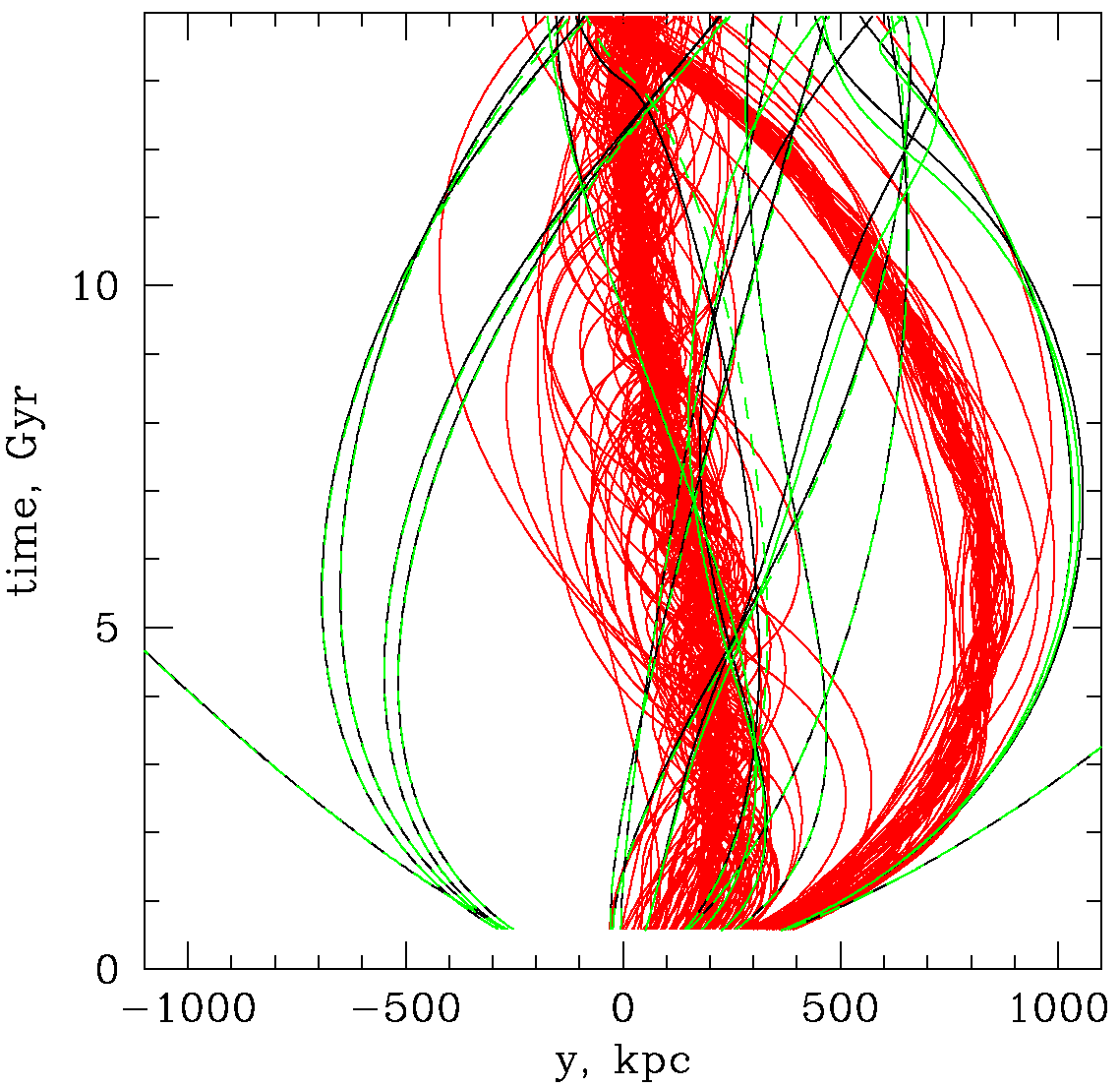} 
\caption{\label{Fig:fwd} \footnotesize Result of numerical integration forward in time from the initial conditions in Figure~1. Red curves are samples of MW and LMC halo particle orbits, solid black curves are PT2 orbits, and green curves are from the computation forward in time. The green curves are solid for the five actor orbits that most differ from PT2, and dashed for the rest.}
\end{center}
\end{figure}

\subsection{Forward Integration Test}\label{Sec:fwdint}

For simplicity the halo reconstruction does not take account of the disturbance to the motions of the actors by the rearrangement of mass within MW and LMC. It would be easy to remove this approximation in a more capable computation. The less complete but even easier test presented here compares the MW halo in reconstruction~(a) to model~(${\rm a}_{\rm f}$), in which actors as well as halo particles move in response to their gravitational interactions. The initial positions at the first two time steps are the same as in Figure~1 for actors and halo particles. The result of leapfrog integration forward in time from there is shown in Figure~\ref{Fig:fwd}. The coordinates are galactic and physical, the origin is the center of mass of the 19 PT2 actors, and the samples of 100 MW and 50 LMC halo particle orbits can be compared to those in the middle panel in Fig.~\ref{Fig:orbits}) (but here all halo particle orbits are plotted in red). The scale is changed to include orbits of more distant actors. The solid black lines are the PT2 actor orbits used in the reconstructions. The green lines are the actor orbits in~(${\rm a}_{\rm f}$). The five~(${\rm a}_{\rm f}$) actor orbits that are most different from PT2 are plotted as solid green lines. The other (${\rm a}_{\rm f}$) actor orbits are plotted as dashed green lines. In many cases these orbits are so little different from PT2 that they appear as dashed alternating green and black lines.  

The mass rearrangement in MW and LMC has the most notable effect on Leo~T, which~(${\rm a}_{\rm f}$) puts 190~kpc from the PT2 position. At PT2 distance 406~kpc from the MW this may be counted as a 50\% error. By this measure the next largest discrepancies are NGC~185 at 40\%, IC~10 at 30\%, and M33 and LGS3 at 20\%.  But since these actors have relatively little mass there is not much effect on the more massive actors. At $z=1$ the MW and LMC halo particle positions in~(a) and~(${\rm a}_{\rm f}$) differ by only $\sim 1$~kpc. At $z=0$ the differences have grown to include a 35~kpc shift in the center of the LMC halo relative to the MW halo. This moves the peak of the DM stream to $L_{\rm MS}\sim 30^\circ$, but the patterns of redshifts and positions as functions of $L_{\rm MS}$ are quite similar to Figure~\ref{Fig:fwd}.

The purpose of this exercise is to explore the reaction of the reconstructed MW halo to the disturbance of the actor orbits caused by the reconstruction of MW and LMC. The dotted lines in Figure~\ref{Fig:vc} show the circular velocity and radial and transverse velocity dispersions in halo~(${\rm a}_{\rm f}$). They are quite close to these measures in MW halo~(a), meaning the structure of the MW halo has been little affected by the disturbances of the actors. In Figure~\ref{Fig:am} the solid and dotted black lines compare the angular momentum direction and magnitude in the~(a) and~(${\rm a}_{\rm f}$) halos. The directions are quite similar, and, although the magnitude of the angular momentum in halo~(${\rm a}_{\rm f}$) is larger than~(a) at $r\sim 100$~kpc, the difference is modest enough to allow the conclusion that the MW halo angular momentum is not seriously affected by the disturbance to the actor orbits caused by the rearrangement of mass in the MW and LMC.

\section{Discussion}\label{Sec:discussion}

The DM stream around the reconstructed LMC halo in Figure~\ref{Fig:MS} resembles the DM stream in the even more schematic model in PT2, and Figure~\ref{Fig:MS} is similar enough to the HI Magellanic Stream to suggest that the overall pattern of the observed stream could have grown out of the gravitational interaction with neighboring galaxies at high redshift and with the MW at low redshift. This is the scenario first considered by Fujimoto  \& Sofue (1976) and Lin \& Lynden-Bell (1977), here applied at high redshift under cosmological initial conditions. It does not seem to be necessary to postulate that the HI Magellanic Stream formed at intermediate redshift by interaction with an actor such as the Small Magellanic Cloud, as in the scenario presented by Besla, Kallivayalil, Hernquist, {\it et al.} (2012), though the interaction Besla {\it et al.} postulate certainly played an important role in shaping the HI bridge between the Large and Small Magellanic Clouds, and perhaps was important also in the growth of other parts of the stream substructure. It may be of interest that the DM streams in the five reconstructions presented here have shorter leading than trailing parts, but the HI in the leading part of the Magellanic Stream is much less prominent relative to the trailing part. The scenario thus requires a postulate, perhaps that ram pressure by a MW corona shortened the leading stream (Moore \& Davis 1994 and references therein), or perhaps that lower density in the leading stream allowed a greater fractional ionization.  

The elongated and lumpy shape of the reconstructed MW halo at redshift $z=3$ (Fig.~\ref{Fig:z3}) is distinctive  enough to suggest that the reconstruction would pass through a state about like this if the initial conditions were more lumpy, and perhaps more realistic, than the smooth distribution in Figure~\ref{Fig:z9}. At $z=2$ the reconstructed DM halo has evolved to about at its present size and shape, though substructure still is visible (Fig.~\ref{Fig:z210}). At $z=1$ the halo looks about as relaxed as it is now. The change of shape and orientation from $z=2$ to $z=0$  may be an artifact of the relatively  small number of halo particles. That  could be checked in a more capable computation with many more halo particles. An even more capable computation could use the initial conditions in Figure~\ref{Fig:orbits} to simulate the behavior of the diffuse baryons and the formation of stars. That may inform us of the meaning of the evidence that the MW disk of stars was in place at $z\sim 2$ and that the bulk of the MW stars ended up largely confined to the plane of the bar and thin and thick disks (Gilmore, Wyse, \& Norris  2002; Kormendy, Drory, Bender, \& Cornell 2010).

In the five reconstructions the angular momentum of the MW halo outside the noisy central part is directed below galactic latitude $b\sim -20^\circ$ (Fig.~\ref{Fig:am}). If the noise in the reconstructions had produced an isotropic probability distribution of angular momentum direction then the probability that all five end up below $b= -20^\circ$ is $((1+\sin b)/2)^5\sim 0.004$. That is, we have a statistically significant case that the  reconstruction puts the MW halo angular momentum in the general direction of the disk (at $b=-90^\circ$). However, we see from the broad scatter of directions in Figure~\ref{Fig:am} that the angular momentum is sensitive to substructure in the initial conditions, and the effect of larger initial substructure remains to be explored. It is to be noted that, although the magnitude of the angular momentum is within the range of values found in numerical cosmological simulations, it is at the extreme low end. This could be because the model does not have much initial subclustering to couple to the gravitational field of the external mass distribution and  transfer angular momentum. This could be checked by commencing the computation at higher redshift where  the mean halo density contrast and its substructure are smaller, and applying substructure to the young protohalo. 

A modest improvement of computer and particle accelerator would allow iterated adjustment of the initial positions of the actors to hold their present positions fixed as some some of the actors are reconstructed. This would avoid the considerations in Section~\ref{Sec:fwdint}. It might then be interesting to reconstruct M\,31 as another DM halo and explore its angular momentum. In a reconstruction with better spatial resolution it may even be possible to compare streaming motion near the center of the reconstructed LMC halo to what is observed (van der Marel \& Kallivayalil 2014). But I expect the more immediately pressing task is to find a more complete description of the distribution of mass external to the reconstructed halos. The large and growing fund of accurate distances of galaxies within 10~Mpc (as in Wu, Tully, Rizzi,  {\it et al.} 2014) together with their redshifts offers the opportunity for a much more complete model than PT2. To be decided is whether that is best done in a representation of the mass distribution by the positions of actors or by an expansion in a basis of functions with resolution adapted to the distance uncertainties.

\end{document}